\documentclass{article}
\usepackage{xypic,latexsym}
\input{amssym.def}
\author{Peter Hines}
\title{Symmetries and transitions of bounded Turing machines}
\begin{document}
\newtheorem{definition}{Definitions}
\newtheorem{theorem}{Theorem}
\newtheorem{prop}[theorem]{Proposition}
\newtheorem{corol}[theorem]{Corollary}
\newtheorem{lemma}[theorem]{Lemma}
\maketitle
\begin{center} Abstract \end{center}
{\em We consider the structures given by repeatedly generalising the definition of finite state automata by symmetry considerations, and constructing analogues of the transition monoid at each step. This approach first gives us non-deterministic automata, then (non-deterministic) two-way automata and bounded Turing machines --- that is, Turing machines where the read / write head is unable to move past the end of the input word.

In the case of two-way automata, the transition monoids generalise to endomorphism monoids in compact closed categories. These use Girard's resolution formula (from the Geometry of Interaction representation of linear logic) to construct the images of singleton words. 

In the case of bounded Turing machines, the transition homomorphism generalises to a monoid homomorphism from ${\Bbb N}$ to a monoid constructed from the union of endomorphism monoids of a compact closed category, together with an appropriate composition. These use Girard's execution formula (also from the Geometry of Interaction representation of linear logic) to construct images of singletons.}

\begin{center} AMS Classification: 68Q05, 86Q70, 18D15 \end{center}

\section{Introduction}
In what follows, we take one of the simplest possible definitions in the theory of computation --- that of a deterministic finite state automaton without specified initial or terminal states --- and repeatedly generalise the definition by symmetry considerations. 
Each successive generalisation leads to a more complicated structure,
albeit already known, if not widely studied. However, the way in which the algebraic models follow the generalisations is of more interest.

This procedure allows us to form algebraic models of each of the structures under consideration, which appear to be, in each case, important tools in answering questions about these structures. It is also of interest to note that the algebraic structures used are similar both to those used by theoretical physicists in their algebraic models of symmetries, and those used by logicians for representing the process of deduction.

\section{Generalising finite state automata}
We take as the starting point of our generalisation process 
the idea of a finite set of functions from a finite set to itself --- that is, we consider a set $Q$, and a set of functions $\Sigma$ from $Q$ to $Q$.
Algebraic models arise when we consider multiple function applications, 
$hgf(a)=hg(b)=h(c)=d$.
We wish keep a track of this process, so we draw the above as 
\[ hgf^{(a)} \ \mapsto \ hg^{(b)}f \ \mapsto \ h^{(c)}gf \ \mapsto \ ^{(d)}hgf .\]
Note that when we draw it this way, we do not forget the function we have just applied. This is because, intuitively, we consider forgetting information to be a computational step.

Consider the following graphical representation of the action of the functions $\{ f,g\}$ on the set $\{ a,b,c\}$
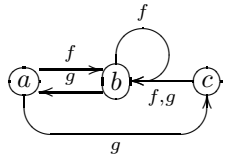
\[ \diagram
 \framed<5pt> a \rrtod_{g} \rto<1ex>^{f} & \framed<5pt> b \lto<1ex>_{g} \tour^{f}               & \framed<5pt> c\lto^{f,g}
\enddiagram
 \]
Using a graphical representation, the action of string of function symbols can be thought of as `following a labelled path through a diagram'. This is, of course, a {\em finite state automaton}; we refer the reader to \cite{HU} for the basic theory.
The set $\Sigma$ is the {\em input alphabet}, and the set $Q$ is the {\em set of states}. As the name suggests, these are both finite. There is also a {\em next state function}, 
$\circ : \Sigma \times Q
\rightarrow Q$ which we have been representing by function application. 

In \cite{DC}, the following model is given:
The set of all words of $\Sigma$ form a monoid, with composition given by concatenation of strings, and the identity element given by the empty word, $\lambda$. This monoid is denoted $\Sigma^*$. Similarly, the set of all functions on $Q$, denoted $PT(Q)$, is also a monoid. Formally, we identify a function $f:Q\rightarrow Q$ with the subset of $Q\times Q$ given by 
$\{ (f(q),q) : q\in Q\}$. The 
composition of this monoid is given by: $(c,a)$ is a member of 
$SR$ if and only if there exists some $b$ with $(c,b)$ a member of $S$ and $(b,a)$ a member of $R$; it is immediate that this composition is associative, and the identity of $PT(Q)$ is the {\em diagonal relation} $I=\{ (q,q): q\in Q\}$.

The {\em transition function} $t:\Sigma^* \rightarrow PT(Q)$ is defined as follows: Let $w$ be a string of function symbols from $\Sigma$. Then the pair $(b,a)$ is a member of $t(w)$ if and only if $w(a)=b$. From the computational interpretation of the function $t$, it is clear that 
$t(u)t(v) = t(uv)$ and $t(\lambda )=I$. Hence $t$ is a monoid homomorphism.
This, together with the fact that we know $t(f)$ for all $f$ in $\Sigma$ allows us to describe $t(w)$ for any $w$ in $\Sigma^*$. This then allows a characterisation of the words of $\Sigma^*$ (i.e. composites of functions) that have the same action on the set $Q$. Hence we can identify words over the input alphabet to form a finite quotient of $\Sigma^*$, called the transition monoid. The homomorphism $t$ uniquely determines, and is uniquely determined by the transitions of the finite state automaton. Also, every monoid homomorphism from a finitely generated free monoid to a monoid of functions on a set determines a finite state automaton.

\subsection{Generalising to non-deterministic automata}
Consider the following transition diagram, $ \diagram
      \framed<5pt> a \tolu^{f,g}  & \framed<5pt> b \lto^{f,g}  \enddiagram$.
As the first step in our `generalising by symmetry' process, we would like to have an operation that reversed all the arrows in this transition diagram, to define another finite state automaton, which would be specified by $\diagram  
  \framed<5pt> a \told_{f,g} \rto_{f,g} & \framed<5pt> b
\enddiagram$. 
However, as it stands, this is not a finite state automaton. We required 
$\Sigma$ to be a set of functions, and in the second diagram $f(a)$ can be either $a$ or $b$, and $g(b)$ is not defined. We replace the requirement that members of $\Sigma$ define functions on the set $Q$ with the more general requirement that they define {\em relations} on $Q$ --- that is, they are arbitrary sets of pairs of elements of $Q$ (an alternative, but equivalent, requirement is that members of $\Sigma$ define functions from $Q$ to the set of all subsets of $Q$, denoted $P(Q)$ ). 

This is the generalisation from {\em deterministic automata} to {\em non-deterministic automata}. With a deterministic automaton, there is exactly one next state, for any $x\in \Sigma$ and $q\in Q$. With a non-deterministic automaton, there is a set of next states (which may be empty). We replace the next state function $\circ : \Sigma \times Q \rightarrow Q$ by the next-set-of-states function $\circ : \Sigma \times Q \rightarrow P(Q)$, where $P(Q)$ denotes the set of all subsets of $Q$.

We can define the transition homomorphism, as before; the transition homomorphism for a non-deterministic automaton is given by a homomorphism from $\Sigma^*$ to $B(Q)$, the monoid of relations on $Q$. Note that every monoid homomorphism from a finitely generated free monoid to a finite monoid of relations uniquely determines a (non-deterministic) finite state automaton. 

The mathematical representation of the `reversing arrows' symmetry is an operation defined on monoids of relations, given by $R\mapsto \overline{R}$, where $(a,b)$ is in $\overline{R}$ if and only if $(b,a)$ is in $R$. Now consider the relation $t(x)$ for some $x\in \Sigma$. The pair $(b,a)$ is in $t(x)$ if there is an arrow from state $a$ to state $b$ labelled by $x$, and the connection between the `reversing arrows' operation, and the symmetry $\overline{(\ )}$ is then apparent.

\section{Generalising to two-way automata}
Recall that we drew transitions of a finite state automaton as a pointer labelled by a state moving from left to right over a string of symbols. This convention was chosen because we were writing function application on the left, which is in turn due to the Arabic basis of our number system. We would like to eliminate the asymmetry given by the arbitrary choice of direction; that is, we would also like to be able to consider transitions of the form
$\  ^{(d)}hgf \ \mapsto \ h^{(c)}gf\ \mapsto \  hg^{(b)}f \ \mapsto \ hgf^{(a)}$, so that
in terms of a pointer moving over words, the {\em direction} of movement is reversed. 

Consider a (non-deterministic) automaton $A$ with state set $Q$, input alphabet $\Sigma$, and next state function $\circ_l: \Sigma \times Q \rightarrow P(Q)$. We eliminate the directional asymmetry by introducing a {\em right-moving next state function} $\circ_r : Q\times \Sigma \rightarrow P(Q)$. 
The interpretation of this is that the pointer labelled by a state may move either left or right on a word (or both --- we allow for the possibility of non-determinism), and change state accordingly. So, the evolution of the system under time proceeds as follows: 
Consider a point in a computational process,
$ \ldots x_{i-2}x_{i-1}\ ^{(a)}\ x_ix_{i+1} \ldots $
The set of next configurations is the union of $\ldots x_{i-2}\ ^{(b)}\ x_{i-1} x_i \ldots $
where $b$ is in $x_{i-1}\circ_l a$ and $ \ldots x_{i-1}x_{i}\ ^{(c)}\ x_{i+1} \ldots $
where $c$ is in $a \circ_r x_i$.

This structure is a {\em two-way automaton}. It is not described in the same way as the usual model, due to J.-C. Birget (see \cite{JCB2}); he labels the states as either left-moving, or right-moving  (or both; the sets of left and right moving states are not assumed to be disjoint) and has a single next-state function. The movement of the pointer is then to the left or the right, depending on whether the current state is left or right moving.

The two different descriptions are not quite the same; consider the computations $\ ^{(q)} x \ \mapsto \ x^{(q')}$ and $x ^{(q)} \ \mapsto \ ^{(q'')}x$.
In Birget's model, $q'$ must be the same as $q''$. Using our model, they may differ.  
However, they can be seen to be equivalent; our model clearly contains Birget's model as a special case (the left and right moving state sets are the domains of $\circ_l $ and $\circ_r$ respectively). Conversely, consider a model defined in our terms, together with a state $q$ that is both left and right moving. We can then `split $q$ into 2 parts', $q_l$ and $q_r$, and 
adjust the next state functions accordingly. Then Birget's single next state function is defined to be the union of our left moving and right moving next state functions (This construction is used in a `non-determinism is (almost) equivalent to determinism' proof of J.-C. Birget for two-way automata \cite{JCB}, and a full proof of the equivalence of the above model with his can be found in \cite{PHD}).

To generalise the transition homomorphism $t$, we need to know when a computation is finished. In the one-way case it was immediate; a computation is finished when it reaches the left of the word of symbols from $\Sigma$. In the two-way case, there are four possibilities, depending on whether the pointer starts on the left or the right, and whether it finishes on the left or the right. Using this basic idea, J.C. Birget constructs algebraic models of two-way automata.
For every word of the input alphabet, he defines four relations on $Q$:
\begin{itemize}
\item 
$[\rightleftharpoons w]$, consisting of  the set of pairs $(q',q)$
for which there exists 
a computation starting in configuration 
$^{(q)} w$ and finishing in configuration $^{(q')}w$, with $q$ rightmoving, and 
$q'$ leftmoving.
\item 
$[-w\rightarrow ]$,  consisting of  the set of pairs $(q',q)$
for which there exists 
a computation starting in configuration 
$^{(q)} w$ and finishing in configuration $w^{(q')}$, with $q$ and $q'$ rightmoving.
\item 
$[\leftarrow w-]$, consisting of  the set of pairs $(q',q)$
for which there exists a computation starting in configuration 
$w^{(q)}$ and finishing in configuration $^{(q')}w$, with $q$ and $q'$ leftmoving.
\item 
$[w\rightleftharpoons ]$, consisting of  the set of pairs $(q',q)$
for which there exists a computation starting in configuration 
$w^{(q)}$ and finishing in configuration $w^{(q')}$, 
with $q$ left-moving and $q'$ rightmoving.
\end{itemize}
These relations, which he calls {\em global transition relations}, 
also feature implicitly in the earlier work, `\cite{JCS}, by J. Shepardson. 
The point of J.-C. Birget's definitions is that, for any two words of $\Sigma^*$, the global transition relation of their composite can be written in terms of their global transition relations, and composition in the monoid of relations. This is as follows:
\begin{theorem} Given a two-way automaton 
${\Bbb A}=(Q=Q_l\cup Q_r,\Sigma ,\circ )$, and 
$u,v\in \Sigma ^*$, then the global transition relations 
of the composite $uv$ are defined in terms of the global transition 
relations of $u$ and $v$, as follows:
\begin{itemize}
\item 
$[-uv\rightarrow ] = [-v\rightarrow ] 
([u\rightleftharpoons ] [\rightleftharpoons v] )^* 
[- u\rightarrow]$, 
\item $[uv\rightleftharpoons ] = [v\rightleftharpoons ] \cup 
[-v\rightarrow ] [u\rightleftharpoons ] ( [\rightleftharpoons v]
[u\rightleftharpoons ] )^*[\leftarrow v-]$,
\item $[\rightleftharpoons uv] = [\rightleftharpoons u] \cup 
[\leftarrow u -][\rightleftharpoons v] 
([u\rightleftharpoons ] [\rightleftharpoons v])^* [-u\rightarrow ]$,
\item $[\leftarrow uv- ] = [\leftarrow u- ] 
( [\rightleftharpoons v][u \rightleftharpoons ])^* [\leftarrow v- ]$.
\end{itemize} 
Where, for a relation $R$, its {\em Kleene star} is defined by 
$R^* = I \cup R \cup R^2 \cup R^3 \cup \ldots$
$\Box$
\end{theorem}

\noindent The algebraic framework for these composition rules is a special case of a construction of Joyal, Street, and Verity, \cite{JSV}, which they refer to as the $Int$ construction\footnote{Although we follow the examples and notation of Joyal, Street and Verity (because they give the example of the category of relations as a special case), the $Int$ construction was first presented (in a different manner) in terms of logical models, in Abramsky and Jagadeesan's `New Foundations for the Geometry of Interaction' \cite{AJ}. The equivalence of the two constructions, together with the intuitive ideas behind the categorical construction - including the close connection with iteration - appeared in Abramsky's `Retracing some paths in process algebra' \cite{SA}.}

In this construction, the category ${\bf Rel}$, that has all sets as objects and relations between sets as arrows (so that the endomorphism monoid of an object $X$ is the monoid $B(X)$ of relations), is `dualised' to form a category ${\bf IntRel}$. This category has 
pairs of sets as objects. An arrow between two objects, say $(X,U)$ and $(Y,V)$, consists of four arrows in the category of relations, as follows:
\[ \diagram 
    X \rto^c \dto_a  & U                  \\
   Y  &   V\uto_d  \lto_b 
\enddiagram  \]
Composition of relations is then giving by a `taking the union over all possible paths' construction. Two squares, as above, are composed vertically, and the union over all possible paths between bottom and top is constructed, as follows:
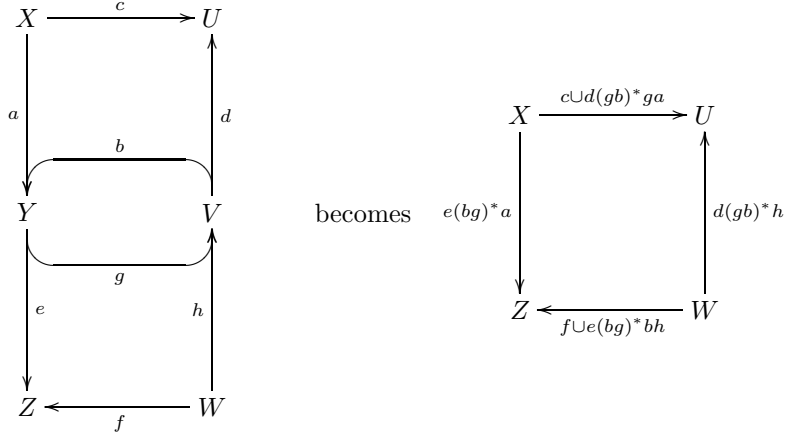
\[ \diagram 
    X \rrto^c \ddto_a  &        & U &                                    &     &   \\ 
                     &  &      & & X \rrto^{c\cup d(gb)^*ga} \ddto_{e(bg)^*a}  &     & U   \\ 
   Y \rrtod_g \ddto^e  &  &   V\uuto_d  \lltou_b &\mbox{ becomes } \ &                                             &     &     \\ 
                       &  &          & &  Z   &     & W \uuto_{d(gb)^*h} \llto^{f\cup e(bg)^*bh} \\             
   Z                   &  & W \llto^f \uuto^h & &                                    &     &  
    \enddiagram  \]
This can also be written more concisely in matrix form as 
\[ \left( \begin{array}{cc} e & f \\
                               g & h \end{array}\right)
\left( \begin{array}{cc} a & b \\
                               c & d \end{array}\right) = 
\left( \begin{array}{cc}
               e(bg)^*a           &   f \cup e(bg)^*bh \\
               c \cup d(gb)^*ga   &   d(gb)^*h  
               \end{array}\right) \]
This then gives context for Birget's equations for two-way automata. 
Define a function $[ \ ]$ from the set of words on the input alphabet, $\Sigma^*$, to the endomorphism monoid of $(Q,Q)$ in ${\bf IntRel}$, by 
\[ [w]=
\left(
         \begin{array}{lr} \ [\leftarrow w-] & \ [\rightleftharpoons w] \\ \ [w\rightleftharpoons ] & \ [-w\rightarrow ]
               \end{array} \right) \]
\begin{theorem}
The composition of $[v]$ and $[u]$ in ${\bf IntRel}$ is given by Birget's equations, so $[vu]=[v][u]$. 
\end{theorem}
{\bf Proof} This follows immediately by comparing the composition of ${\bf IntRel}$ with Birget's equations.
$\Box$ \\

\noindent
The above construction shows that the map $[\ ]$ is a semigroup homomorphism - it is not a monoid homomorphism; the empty word $\lambda$ is mapped to the idempotent 
$\left( \begin{array}{cc}0 & 1_{Q_l} \\
1_{Q_r} & 0 
\end{array} \right) $. 
This makes the image of $\Sigma^*$ under $[\ ]$ what \cite{JH1} refers to as a {\em local submonoid} of the endomorphism monoid of $(Q,Q)$.

Unlike the one-way case, the value of $[\ ]$ for words of length 1 (that is, members of $\Sigma$) is not immediate from the definition. These can be found using a tool developed by J.-Y. Girard for linear logic, in \cite{GOI1}.

\subsection{Girard's resolution formula, and two-way automata}
In \cite{GOI1,GOI2,GOI3}, J.-Y. Girard constructed a novel series of representations of linear logic (see \cite{LL} for an introduction to this logical system), called the {\em Geometry of Interaction}. This representation was in terms of matrices over monoids of relations. A common feature of all these was the {\em resolution formula}. This was defined in terms of matrices of relations (where, in the composition of matrices of relations, multiplication is interpreted by monoid composition, and addition is interpreted by union), to be
\[ Res (U,\sigma )= \pi U(1-\sigma U)^{-1}\pi = \pi U(\sigma U)^* \pi ,\]
where $\pi$ is 
an idempotent given by a $\{(a,a) :a\in A\}$, for some $A\subseteq {\Bbb N}$. 
The matrix of relations $\sigma$ was also assumed to be (in the simplest case) the anti-diagonal matrix
$\left( 
         \begin{array}{cc}
           0 & 1 \\ 1 & 0
         \end{array} \right)$,
where 1 and 0 are, respectively, the identity and nowhere-defined relations on $B(\Bbb N)$.
The connection of this formula with the theory of two-way automata is then as follows:
\begin{theorem} Let
${\Bbb A} = (Q,\Sigma ,\circ_l ,\circ_r )$ be 
a two-way automaton, and consider $x\in \Sigma$.
Define 
\[ Q_l= dom(\circ_l ) \ ,\ Q_r=dom(\circ_r) \ ,\ j=t_l(x) \ ,\ k=t_r(x) \] 
where 
$t_l$  and $t_r$ are the transition homomorphisms for $(Q,\Sigma,\circ_l)$ and $(Q\,\Sigma ,\circ_r)$ respectively.
A matrix consisting of  
the global transition relations of $x$ is given by 
the following version of Girard's resolution formula over the monoid of relations
on $Q$ :
\[ Res(U,\sigma ) = 
\left( 
         \begin{array}{cc}
           1_{Q_l} & 0 \\ 0 & 1_{Q_r}
         \end{array} \right) \left(
\begin{array}{cc}
             k & 0 \\ 0  &  j  
       \end{array} \right) 
\left[ 
\left( \begin{array}{cc} 0 & 1 \\ 1 & 0 \end{array} \right)
\left(
\begin{array}{cc}
             k & 0 \\ 0  &  j  
       \end{array} \right) 
\right]^*
\left( 
         \begin{array}{cc}
           1_{Q_l} & 0 \\ 0 & 1_{Q_r}
         \end{array} \right) 
\]
\end{theorem}
{\bf Proof} First note that a direct calculation will give  
\[ Res (U,\sigma ) = 
\left( \begin{array}{cc}
               1_{Q_l} k(jk)^* 1_{Q_l} & 1_{Q_l}(kj)(kj)^* 1_{Q_r} \\
               1_{Q_r}(jk)(jk)^*1_{Q_l}    & 1_{Q_r}j(kj)^*1_{Q_r} 
          \end{array} \right) .\]
Conversely, the transitions of a two-way automaton on the symbol $x$ can be represented as follows: 
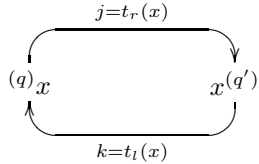
\[ \diagram 
        ^{(q)} x  \rrtou^{j=t_r(x)} & &  x^{(q')} \lltod^{k=t_l(x)}             
   \enddiagram \]
All possible transitions from configurations 
of the form $^{(q)}x$ to configurations of
the form $^{(q')}x$ are given by $kj$, $(kj)^2$, $(kj)^3$, $\ldots$ 
Hence, $[\rightleftharpoons x]$ is the intersection of $(kj)(kj)^*$ with 
$Q_l\times Q_r$, and so 
$[\rightleftharpoons x] = 1_{Q_l}(kj)(kj)^*1_{Q_r}$.
This is the top right entry of $Res (U,\sigma )$, as required. Similar considerations will give 
the other three global transition relations, 
$[\leftarrow x-]$, $[x\rightleftharpoons ]$, and 
$[-x\rightarrow ]$ as the top left, bottom left, and bottom right entries respectively. $\Box$\\

\subsection{Interpretation of the `reversing directions' symmetry}
The category ${\bf IntRel}$ is a compact closed category --- the point of the $Int$ operator is a canonical construction of compact closed categories. 
These are special cases of symmetric monoidal categories (we refer to \cite{MCL} for the theory of symmetric monoidal categories, and their coherence equations), where for every object $A$, there exists another object $A^\vee$, called its {\em (left) dual}, satisfying: there exists an arrow from $I$ to $A \otimes A^\vee$, and an arrow from $A^\vee \otimes A$ to $I$ (together with natural coherence conditions which were mathematically analysed in \cite{KL}.)\footnote{As could be deduced from this very suggestive notation, compact closed categories have been heavily used in theoretical physics --- see \cite{JB} for more details.} 

These definitions imply the existence of {\em duals on arrows}, so that, for any arrow $f:A\rightarrow B$, there exists its dual $f^\vee : B^\vee \rightarrow A^\vee$. In the category ${\bf IntRel}$, the dual on objects is $(X,U)^\vee = (U,X)$ and the dual on arrows is given by  
 \[ \left( \begin{array}{cc} a & b \\ c & d \end{array} \right)^\vee = 
\left( \begin{array}{cc} d & c \\ b & a \end{array}\right) \]
The dual on arrows of this category then has the simple interpretation of interchanging the left and the right moving parts of a two-way automaton (i.e. taking $\circ_r$ as the left-moving next state function, and dually for $\circ_r$). Hence, if $[w]$ is the global transition relation of $w$ for the automaton ${\Bbb A}$, then $[w]^\vee$ is the global transition relation of $w$ for the automaton given by swapping the left and right moving parts of ${\Bbb A}$.

\subsection{Summary of algebraic models of two-way automata}
It is worthwhile just to summarise the differences between algebraic models of one-way and two-way 
automata. They are both functions from the set of all words on the input alphabet;
in the one-way case, a monoid homomorphism to the monoid of relations on the set of states $Q$ and 
in the two-way case, a homomorphism to the endomorphism monoid of $(Q,Q)$ in the compact closed category ${\bf IntRel}$. 
What allows us to calculate them explicitly is that we know values for members of $\Sigma$;
in the one-way case, these are immediate from the definition, and 
in the two-way case, these are given by Girard's Resolution formula.

Note that a one-way (non-deterministic) automaton is uniquely specified by a monoid homomorphism, as above. However, a two-way automaton is not uniquely specified by such a homomorphism; the images of members of $\Sigma$ are of a very special form.

The definition of a one-way automaton has one symmetry, given by reversing the direction of arrows in a transition diagram. 
The mathematical representation of this is the $\overline{(\ )}$ operation in the category of relations.
The definition of a two-way automaton also has the symmetry given by reversing the direction of movement. The mathematical representation of this is the dual operator in a compact closed category. 
The `reversing arrows' operation on the category of relations generalises directly (by the $Int$ construction) to the compact closed category ${\bf IntRel}$. This makes ${\bf IntRel}$ what J. Baez refers to as a `compact closed category with duality'; see \cite{JB} for more details.

\section{Generalising to bounded Turing machines}
The next observation that allows us to continue our generalisation process is that in the definition of a finite state automaton, the set of states has a distinguished r\^ole. Given a state and a symbol, we have a function that gives us another state. So, what we require in order to make this definition symmetric with respect to states and alphabets is another function (or two new functions in the two-way case) that takes a state and a symbol, and gives us a new symbol. We first consider how this generalises the definition of one-way finite state automata, and then extend to the two-way case.

\subsection{Generalising one-way automata to Mealy machines}
Given a (non-deterministic) automaton $A=(Q,\Sigma ,\circ :\Sigma\times Q\rightarrow P(Q) )$, 
we wish to make this definition
symmetrical between the state set, and the input alphabet. To do this, we 
introduce the {\em output function} $*: \Sigma \times Q \rightarrow P( \Sigma )$.
The interpretation is that each state change has an output
associated with it, so an input word of length $n$ will give a set of output words of
length $n$. Consider the following example, with state set $\{ n,c\}$ and alphabet $\{ 0,1\}$, where $\circ$ and $*$ are specified by 
$
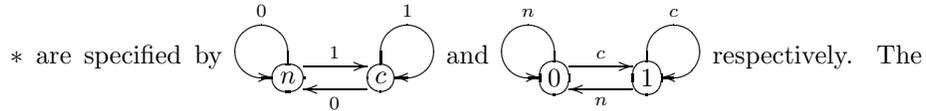
\diagram 
 \framed<5pt> n \toul_{0} \rto<1ex>^{1}  & \framed<5pt> c \tour^{1} \lto<1ex>^{0} \enddiagram$ and $\diagram 
 \framed<5pt> 0 \toul_{n} \rto<1ex>^{c}  & \framed<5pt> 1 \tour^{c} \lto<1ex>^{n}
\enddiagram $
respectively. The computation of $R$  on the input $101101$ starting in state $n$ is then 
\[ \begin{array}{ccccccccc}  
                     1 & 0 & 1 & 1 & 0  & 1 & \ ^{(n)} &  & \\
                     1 & 0 & 1 & 1 & 0 & \ ^{(c)} & 1   &  & \mbox{ output:} 0 \\
                     1 & 0 & 1 & 1 & \ ^{(n)} & 0   & 1   &  & \mbox{ output:} 1  \\
                     1 & 0 & 1 & \ ^{(c)} & 1 & 0   & 1   &  & \mbox{ output:} 0 \\
                     1 & 0 & \ ^{(c)} & 1 & 1 & 0   & 1   &  & \mbox{ output:} 1 \\  
                     1 & \ ^{(n)} & 0 & 1 & 1 & 0   & 1   &  & \mbox{ output:} 1 \\  
                    ^{(c)}\ & 1 & 0 & 1 & 1 & 0   & 1   &  & \mbox{ output:} 0    
      \end{array} \]
This gives the output word $011010$, and ends in state $c$.
As can be seen from this example, the action of this machine is to left-shift a
binary string (i.e. multiply by two), and record if there is
overflow (where $c=$`carry bit set' and $n=$`no carry').
Note that the above pair of transition diagrams define two `automata with output', one with state set $Q$ and input/output alphabet $\Sigma$, and the other with state set $\Sigma$, and input/output alphabet $Q$. In fact, the above example is equivalent to itself under this symmetry.

Automata with output are know as {\em Mealy Machines}, or {\em 
Mealy-type automata} (see \cite{DC} for the general theory, and the related {\em Moore machines}, which are automata with output, where every state has an associated output, rather than every transition). The examples we have been considering above are the special cases where the input alphabet is the same as the output alphabet. 

\subsection{Generalising two-way automata to bounded Turing machines}
In the above sections, we took a model of applying finite functions to a finite set and considered in which ways the definition was asymmetric. We first generalised to non-determinism by requiring that a `reversing arrows' operation be well-defined, and then demonstrated how we could either generalise to two-way automata (by left / right movement symmetry), or to Mealy machines (by state / alphabet symmetry).

We now wish to generalise by both at the same time, to get a model of a computing device that can move left and right, and overwrite its input as it goes.
This will give us a model of computation which we refer to as a {\em bounded Turing machine}, which is a Turing machine where the read/write head is unable to move over the end-marker of the tape.

In the following section, we use a significantly different model of Turing machines to the usual, as found in, for example \cite{DC}. The main difference is that we require the pointer to be positioned between cells on a tape, rather than pointing at a cell on a tape, as is usual (see \cite{DC} for an explanation of the usual description). However, our description follows inevitably from the generalisation process, and we feel justified in this approach, because the resulting model is algebraically tractable.\\

\noindent {\bf Definition}
We define (our model of) a bounded Turing machine to be specified by 
\begin{itemize} 
\item a {\em state set}, 
\item $Q$, a {\em set of symbols} $\Sigma$, 
\item two {\em next state functions}, 
$\circ_l :  \Sigma\times Q \rightarrow P(Q)$ and $\circ_r : Q\times \Sigma \rightarrow P(Q)$ that specify the left and right moving state changes, 
\item two {\em rewrite functions}, $*_l : \Sigma \times Q \rightarrow P(\Sigma )$ and $*_r : Q \times \Sigma \rightarrow P(\Sigma )$, that specify the left and right moving rewrites.
\end{itemize}
The action of this machine is as follows:\\
Consider a point in a computational process (a {\em configuration}, or  {\em instantaneous description}),
\[ \ldots x_{i-2}x_{i-1}\ ^{(a)}\ x_ix_{i+1} \ldots \] 
Then the set of next possible configurations is the union of 
all configurations of the form 
\[  \ldots x_{i-2}\ ^{(b)}\ y x_i \ldots \ 
\mbox{ where } b \in x_{i-1}\circ_l a\ ,\ y\in x_{i-1} *_l a \]
and all configurations of the form
\[ \ldots x_{i-2}x_{i-1}\ \ z\ ^{(c)}x_{i+1} \ldots \ 
\mbox{ where } c \in a*_r x_i \ ,\ z\in a\circ_r x_i \]

\subsection{Algebraic models of bounded Turing machines}
We now construct algebraic models of the computations of bounded Turing machines. We cannot directly copy Birget's formul\ae\ in an attempt to construct algebraic models of bounded Turing machines - it makes no sense to write a composite like $[w \rightleftharpoons ][\leftarrow w -]$ in the context of bounded Turing machines;  once the pointer has passed over the word $w$ (that is the $[\leftarrow w -]$ part), then another word has been written on the tape, so we cannot consider $[\rightleftharpoons w]$ as a next step of the computation.

Another, less serious, objection to copying Birget's model directly is the restriction of members of global transition relations to left-moving or right-moving states (recall that $[\leftarrow w -]$ was defined to be the set of pairs $(b,a)$ satisfying a computational condition, subject to the restriction that $a$ and $b$ were both left-moving). This is because, in our formalisation, it is the next state function that is split into two, rather than the set of states.

Because of the above points, we pair words and states, and consider relations consisting of them, as follows:

\noindent
{\bf Definition}\\
For a bounded Turing machine ${\Bbb T}$, and a natural number $n$, we define $\langle \leftarrow n - \rangle$ to be the relation on $\Sigma^n \times Q$ given by 
\[ 
\langle \leftarrow n - \rangle = \left\{ 
\left( \left[
\begin{array}{c} b \\ v \end{array}
\right] , \left[
\begin{array}{c} a \\ u \end{array}
\right] \right) \right\}
\]
where there exists a computation of ${\Bbb T}$ starting in configuration 
$u^a$, and leading to configuration $\ ^bv$, where words $u$ and, by implication, $v$ are of length $n$. We make the dual definition for $\langle - n \rightarrow \rangle$. 

We also require the other two possibilities: we define 
\[ \langle \rightleftharpoons  n \rangle = \left\{ \left( \left[
\begin{array}{c} b \\ v \end{array}
\right] , \left[
\begin{array}{c} a \\ u \end{array}
\right] \right) \right\} \]
where exists a computation of ${\Bbb T}$ starting in configuration 
$\ ^au$, and leading to configuration $\ ^bv$', where words $u$ and $v$ are again of length $n$. Of course, we make the dual definition for $\langle n \rightleftharpoons \rangle$.
\\

\noindent
We call these the {\em computation relations} for the bounded Turing machine ${\Bbb T}$. Note that these relations give, for a bounded Turing machine ${\Bbb T}$ and a fixed tape length $n$,  {\em all possible} computations of ${\Bbb T}$ that start and end at a boundary of the tape.
What is then required is a description of how algebraically tractable these objects are. The first point is that we can write the computation relations in terms of themselves. This is not as trivial as it first appears; consider the following:
\begin{prop}
Let ${\Bbb T}$ be a bounded Turing machine with state set $Q$ and alphabet $\Sigma$. Then the computation relations for words of length $n$ 
satisfy the following:
\begin{itemize}
\item $\langle \leftarrow n - \rangle = 
  \langle \rightleftharpoons n \rangle 
\langle \leftarrow n - \rangle 
( \langle - n \rightarrow \rangle  \langle \leftarrow n -\rangle )^* 
\langle n \rightleftharpoons \rangle$
\item $\langle - n \rightarrow \rangle =
\langle \rightleftharpoons n \rangle \langle - n \rightarrow \rangle 
( \langle \leftarrow n - \rangle \langle - n \rightarrow \rangle )^*
\langle n\rightleftharpoons \rangle$
\item $\langle \rightleftharpoons n\rangle = 
\langle \rightleftharpoons n \rangle 
(\langle \leftarrow n - \rangle \langle - n \rightarrow \rangle )^* 
\langle \rightleftharpoons n \rangle$
\item $\langle  n \rightleftharpoons \rangle = 
\langle n \rightleftharpoons \rangle 
(\langle - n \rightarrow \rangle \langle \leftarrow n - \rangle )^* 
\langle n \rightleftharpoons \rangle$
\end{itemize}
\end{prop}
{\bf Proof} 
Consider 
\[ \left( \left[ \begin{array}{c} q' \\ v \end{array} \right] , 
         \left[ \begin{array}{c} q \\ u \end{array} \right] 
         \right) \in \langle \leftarrow n -\rangle 
\ ,\  
   \left( 
         \left[ \begin{array}{c} q'' \\ w \end{array}
         \right] ,
         \left[ \begin{array}{c} q' \\ v \end{array} 
         \right] 
   \right) \in \langle n \rightleftharpoons
\rangle \] 
By definition of compostion in the category of relations, 
\[ \left( 
      \left[ \begin{array}{c} q'' \\ w\end{array} 
      \right] ,
      \left[ \begin{array}{c} q \\ u \end{array} 
      \right] \right) \in 
\langle n \rightleftharpoons\rangle 
\langle \leftarrow n -\rangle \] 
However, by definition of $\langle \leftarrow n -\rangle$ as the set of 
{\em all} right to left computations, 
\[ \left( \left[ \begin{array}{c} q'' \\ w \end{array} \right] ,
\left[ \begin{array}{c} q \\ u \end{array} \right] \right) \in \langle \leftarrow n -\rangle \]
Hence 
$
\langle \leftarrow n -\rangle
\langle n \rightleftharpoons \rangle
\subseteq \langle \leftarrow n -\rangle$.
However, as we are not restricting by left-moving or right-moving states, the identity relation $I$ is contained in $\langle n \rightleftharpoons \rangle$. Therefore, we can also write 
\[ \langle \leftarrow n -\rangle
\langle n \rightleftharpoons \rangle
= \langle \leftarrow n -\rangle \]
Similar reasoning applies to right to left and left to right computations, so that \[ \langle \leftarrow n -\rangle
\subseteq ( \langle - n \rightarrow \rangle  \langle \leftarrow n -\rangle )^* \]
and as the identity relation is a member of $R^*$ for any relation $R$, 
$\langle \leftarrow n -\rangle
= \langle \leftarrow n - \rangle ( \langle - n \rightarrow \rangle  \langle \leftarrow n -\rangle )^*$.
Finally, dual reasoning to the first identity will give us $\langle \leftarrow n -\rangle = \langle \rightleftharpoons n \rangle \langle \leftarrow n - \rangle$.
Putting these together will give us the identity
$\langle \leftarrow n - \rangle = 
  \langle \rightleftharpoons n \rangle 
\langle \leftarrow n - \rangle 
( \langle - n \rightarrow \rangle  \langle \leftarrow n -\rangle )^* 
\langle n \rightleftharpoons \rangle$.

The symmetry of the bounded Turing machine with respect to left / right movement will then give us the dual of this for free, so that $\langle - n \rightarrow \rangle =
\langle \rightleftharpoons n \rangle \langle - n \rightarrow \rangle 
( \langle \leftarrow n - \rangle \langle - n \rightarrow \rangle )^*
\langle n\rightleftharpoons \rangle$.
This gives us the left to right and the right to left movements in terms of themselves and the other computation relations. 

It can then be shown by similar methods that 
$\langle \rightleftharpoons n\rangle = 
\langle \rightleftharpoons n \rangle 
(\langle \leftarrow n - \rangle \langle - n \rightarrow \rangle )^* 
\langle \rightleftharpoons n \rangle$, and duality with respect to left / right movement gives us 
$\langle  n \rightleftharpoons \rangle = 
\langle n \rightleftharpoons \rangle 
(\langle - n \rightarrow \rangle \langle \leftarrow n - \rangle )^* 
\langle n \rightleftharpoons \rangle$.
This then completes our proof. $\Box$ \\

\noindent
{\bf Definition}\\
As in the two-way automaton case, we can use these four computation relations to construct a member of the endomorphism monoid of $(Q\times \Sigma^n,Q\times \Sigma^n)$ in {\bf IntRel} as follows:
\[ \langle n \rangle = \ \ 
\diagram 
    Q\times \Sigma^n \rto^{\langle\rightleftharpoons n\rangle} \dto_{\langle \leftarrow n - \rangle} &    Q\times \Sigma^n                 \\
   Q\times \Sigma^n  &   Q\times \Sigma^n \uto_{\langle - n \rightarrow \rangle}  \lto_{\langle n \rightleftharpoons \rangle }  \\
\enddiagram  \]
We also make the formal definition that $\langle 0 \rangle$ is the identity at 
$(Q\times \{ \lambda\} ,Q\times \{ \lambda\} )$.
We refer to $\langle n\rangle$ as the {\em computation relation} of the machine ${\Bbb T}$ on words of length $n$.

As expected, we can write Theorem 4 as another `summing over all paths' construction; this demonstrates a symmetry in the the definition of ${\bf IntRel}$. In the definition of the composition of the endomorphism monoid of $(X,X)$ in {\bf IntRel}, there was no a priori reason to give vertical composition special status. For squares of this form, horizontal composition and summing over all possible paths is also a possibility. That is,  
the following is an alternative (and equivalent) definition of composition:
\[ \diagram 
    X \rrto^c \ddto_a  &  &    X \rrto^t \ddtor^r & &  X                 &   &   &  X \rrto^{t(dr)^*c} \ddto_{a\cup b(rd)^*rc}  &     & X   \\
                       &  &                       & &                    & = &   &                                              &     &     \\ 
    X                  &  &    X \llto^b \uutol^d & &  X \uuto^u \llto^s &   &   &  X                                           &     & X \uuto_{u\cup t(dr)^*ds} \llto^{b(rd)^*s}        
    \enddiagram  \]
We now have two distinct compositions on endomorphism monoids of compact closed categories. Of course, taking the union over the set of all possible paths between points is independent of the way the paths were constructed, so given members $a,b,c,d$ of the endomorphism monoid of $(X,X)$, then $\left( \begin{array}{c} a \\ c\end{array} \right)
   \left( \begin{array}{c} b \\ d\end{array} \right) \ = \ 
   \begin{array}{c} (a \ \ b) \\ (c \ \ d) \end{array}$
So, if we (temporarily) write compositions linearly, and denote the vertical and horizontal compositions by $\circ$ and $\cdot$ respectively, then the above becomes
$(a\cdot b) \circ (c\cdot d) = (a\circ c) \cdot (b \circ d)$.
This is the {\em interchange law}, and is a one-dimensional way of representing a two-dimensional equation\footnote{We refer to \cite{RB} for examples and applications to topology, and \cite{JB} for applications to quantum field theories - in particular, using compact closed categories.}. Proposition 4 then states that 
\begin{theorem} Computation relations of a bounded Turing machine are idempotent with respect to the horizontal composition.
\end{theorem}

\subsection{Calculating computation relations}
In the above section we derived, in terms of compact closed categories, formul\ae\ that computation relations must satisfy; however, we have not yet found any way of constructing computation relations, or of writing the computation relations of words in terms of computation relations of shorter words.  We first need 
these relations for words of length one. This can be done using the same tools as the calculation of global transition relations for two-way automata: 
\begin{theorem}
The computation of a bounded Turing machine on words of length 1 can be calculated in terms of a two-way automaton computation.
\end{theorem}
{\bf Proof}
We define a two-way automaton that has a single character, say $1$, for its input alphabet, has $Q \times \Sigma$ as its set of states, and next state functions $\cdot_l$ and $\cdot_r$ given by $1 \cdot_l (x,q) = (x*_l q , x\circ_l q )$ and $(x,q) \cdot_r 1 = (q *_r x , q \circ_r x)$.
It can be seen from the construction of this two-way automaton that if the pair 
$((y,p),(x,q))$ is in $[\leftarrow 1 -]$, then there is a computation of our bounded Turing machine that starts in the configuration $x^q$ and goes to in the configuration $\ ^py$.
Similar results apply for $[- 1 \rightarrow]$, $[\rightleftharpoons 1 ]$, and 
$[1 \rightleftharpoons]$. So we can calculate the computation relations for words of length 1 using a version of Girard's resolution formula (see the discussion of algebraic models of two-way automata) that does not use the idempotent $\pi$ (since we do not restrict to `leftmoving' or `rightmoving' states)\footnote{This variation on the resolution formula is also used in the Geometry of Interaction series of papers, where it is called the {\em execution formula}. However, a discussion of the similarities will take us too far from our original aim, so we refer to \cite{GOI2}.}. 
$\Box$ \\

\noindent
What the above trick does, intuitively, is to `push all the computation onto the set of states and the next state function'. However, we cannot immediately reconstruct the behaviour of a bounded Turing machine using Birget's composition. The above assumes that we are considering words of length 1, and any composition using Birget's formul\ae\ will at most give us more information about the behavior of words of length 1.

To calculate the computation relation of $\langle m+n\rangle$ in terms of $\langle n\rangle$ and $\langle m\rangle$, note that in general, $\langle m\rangle$ and $\langle n\rangle$ are members of two different monoids, and we require a result that is in a third monoid. For clarity, we denote the set $\Sigma^n\times Q$ by $C_n$, and refer to the monoid $B(C_n)$

What we require is functions that take 
relations in $B(C_n)$ to relations in $B(C_{m+n})$, for all $n\in {\Bbb N}$. We define two functions, as follows:\\
$r_{m}:B(C_n) \rightarrow B(C_{m+n})$ is defined by
\[ r_{m} (S) = \left\{ \left(\left [ \begin{array}{c} q' \\ wv \end{array}\right] , 
 \left[ \begin{array}{c} q \\ wu \end{array} \right] \right) \ : \  
 \left( \left[ \begin{array}{c} q' \\ v \end{array} \right]  , 
 \left[ \begin{array}{c} q \\ u \end{array} \right] \right) \  \in S  \ ,\ w\in \Sigma^m \right\} \]
and dually, $l_{n} : B(C_m)\rightarrow B(C_{m+n})$ is defined for all $m\in {\Bbb N}$ by
\[ l_{n} (T) = \left\{ \left(\left [ \begin{array}{c} q' \\ vw \end{array}\right] , 
 \left[ \begin{array}{c} q \\ uw \end{array}\right] \right)\ : \  
 \left( \left[ \begin{array}{c} q' \\ v \end{array}\right] , 
 \left[ \begin{array}{c} q \\ u \end{array}\right]\right)\  \in T  \ ,\ w\in \Sigma^n \right\} \]
\begin{lemma} $ $ \\
{\bf (i)}
The functions $l_{n}:C_a\rightarrow C_{a+n}$ and $r_{m}:C_a\rightarrow C_{m+a}$ are homomorphisms for all $a\in {\Bbb N}$. \\
{\bf (ii)}
Let $Y$ be a relation in $B(C_n)$ and let $X$ be a relation in $B(C_m)$. Then 
$l_m(Y)r_n(X) =$
\[  \left\{ \left(  \left[ \begin{array}{c} q'' \\ zv \end{array}\right]\ , 
                               \left[ \begin{array}{c} q \\ yu \end{array}\right] \right) \ : \ 
                       \left(  \left[ \begin{array}{c} q'' \\ z \end{array}\right]\ , 
                              \left[ \begin{array}{c} q' \\ y \end{array}\right] \right) \in Y \ ,\ 
                       \left(  \left[ \begin{array}{c} q' \\ v \end{array}\right]\ , 
                               \left[ \begin{array}{c} q \\ u \end{array}\right] \right) \in X \right\} .\]
\end{lemma}
{\bf Proof} Both the above results follow directly from the definition of composition in monoids of relations, and from the definitions of $l_{m}$ and $r_{n}$.
$\Box$ \\

\noindent
Now consider two right-to-left computations of a bounded Turing machine; the first on a word of length $m$, and the second on a word of length $n$. Assume the first one takes configuration $y^{q'}$ to configuration $\ ^{q''}z$. Similarly, the second takes configuration $u^q$, to configuration $\ ^{q'}v$. Then it is immediate that  there exists a computation of our bounded Turing machine, on a word of length $m+n$ that from configuration $yu^q$ to configuration $\ ^{q''}zv$. 
We draw this as 
\[ \left. \begin{array}{ccc}
   ^{q'} \ y_m \ldots y_1\   &   \Leftarrow  & u_n \ldots u_1\   ^q \\
 ^{q''} \    z_m \ldots z_1    &  \Leftarrow   &   v_n \ldots v_1 \ ^{q'} 
\end{array}  \right\} \begin{array}{ccc} 
^{q''}  \   z_m \ldots z_1   v_n \ldots v_1  & \Leftarrow & 
y_m \ldots y_1   u_n \ldots u_1 \  ^q \end{array}
 \]
Comparing this with Lemma 11 above then gives the computational interpretation of the $l$ and $r$ functions. 

\noindent
We can apply $l_{n}$ and $r_{m}$ to the 4-tuples of relations of $B(C_n)$ that make up members of $IntRel (C_n,C_n)$ to form two functions 
$R_{m} : IntRel (C_n,C_n)\rightarrow IntRel (C_{m+n},C_{m+n})$
and 
$L_{n} :IntRel (C_m,C_m)\rightarrow IntRel (C_{m+n},C_{m+n})$, in a natural way, as follows: 
\[ R_{m} \left( \diagram 
    C_n \rto^{\langle\rightleftharpoons n\rangle} \dto_{\langle \leftarrow w - \rangle} &    C_n                 \\
   C_n  &   C_n \uto_{\langle - w \rightarrow \rangle}  \lto_{\langle w \rightleftharpoons \rangle }  \\
\enddiagram  \right) = \diagram 
    C_{m+n} \rrto^{r_{m} (\langle\rightleftharpoons n\rangle )} \ddto_{r_{m} (\langle \leftarrow w - \rangle )} &   & C_{m+n}                 \\
                &                           &                 \\
   C_{m+n}  &  & C_{m+n} \uuto_{r_{m} (\langle - w \rightarrow \rangle )}  \llto_{r_{m} (\langle w \rightleftharpoons \rangle )}  \\
\enddiagram  \]
The definition for $L_{n}$ is similar.
The main result of this paper then follows from these two definitions:
\begin{theorem} Consider a bounded Turing machine {\Bbb T}, and let its computation relations for words of length $m$ and $n$ be given by 
$\langle m\rangle$ and $\langle n\rangle$ respectively. Then its computation relations for words of length $m+n$ are given by 
\[ \langle m+n\rangle = R_{m} (\langle n \rangle )\circ L_{n}(\langle m \rangle ) \]
where $\circ$ denotes the vertical composition in the monoid of $(C_{m+n},C_{m+n})$ in ${\bf IntRel}$.
\end{theorem}
{\bf Proof} {\em In the following proof, we denote $r_{m}(\langle \leftarrow  n - \rangle )$ by $r(\langle \leftarrow n- \rangle )$, for simplicity, and similarly for $l$.}

\noindent
We first consider the case of $\langle \leftarrow m+n - \rangle$. By the above example, 
\[ l (\langle \leftarrow m - \rangle )r( \langle \leftarrow n - \rangle ) \subseteq  \langle \leftarrow m+n - \rangle \]
However, we can say more; consider a bounded Turing machine computation on a word of length $m+n$, and mark the intersection between cell $m$ and cell $m+1$ (counting from the left) by the symbol $@$.
So, our cells are numbered 
\[ 1\ 2\ 3\ \ldots m\ ^@ \ m+1 \ \ldots\ m+n \]
Now consider a computation of ${\Bbb T}$ on this tape that starts on the right and finishes on the left, and count the number of times the read / write head passes through the point $@$ (The {\em crossing number}). Also assume that this computation starts in state $p$ and ends in state $q$, and takes an input word $yu$ to an output word $zv$. By the above example, if the read / write head passes through the point $@$ once, then 
\[ \left(  \left[ \begin{array}{c}  q \\ zv \end{array} \right] , 
    \left[ \begin{array}{c}  p \\ yu \end{array} \right] \right) \ \in l( \langle \leftarrow m - \rangle ) r( \langle \leftarrow n - \rangle ) .\]
Similarly, using lemma7, if the read head passes through the point $@$ three times (clearly, it cannot pass through an {\em even} number of times on its way from the right hand side to the left hand side) then we must have that 
\[ \left( \left[ \begin{array}{c}  q \\ zv \end{array} \right] , 
    \left[ \begin{array}{c}  p \\ yu \end{array} \right]\right)  \ \in l (\langle \leftarrow m - \rangle ) 
r (\langle \rightleftharpoons n \rangle ) l( \langle m\rightleftharpoons \rangle )
 r( \langle \leftarrow n - \rangle ) .\]
Then if the read head passes through the point $@$ five times 
\[ \left( \left[ \begin{array}{c}  q \\ zv \end{array} \right] , 
    \left[ \begin{array}{c}  p \\ yu \end{array} \right] \right)  \ \in l (\langle \leftarrow m - \rangle ) 
(r ( \langle \rightleftharpoons n \rangle ) l(  \langle m\rightleftharpoons \rangle ) )^2
r( \langle \leftarrow n - \rangle ) .\]
In general, whatever the crossing number of the point $@$, 
\[ \left( \left[ \begin{array}{c}  q \\ zv \end{array} \right] , 
    \left[ \begin{array}{c}  p \\ yu \end{array} \right] \right) \ \in l (\langle \leftarrow m - \rangle ) 
(r (\langle \rightleftharpoons n \rangle ) l ( \langle m\rightleftharpoons \rangle ) )^i
r ( \langle \leftarrow n - \rangle ) \]
for some natural number $i$.
Therefore, 
\[ \langle \leftarrow m+n - \rangle =l ( \langle \leftarrow m - \rangle )
( r( \langle \rightleftharpoons n \rangle ) l ( \langle m\rightleftharpoons \rangle ) )^*
r ( \langle \leftarrow n - \rangle ) \]

The left / right moving symmetry gives us the dual of this with no further calculation, so  
\[ \langle -m+n \rightarrow \rangle = r ( \langle -n\rightarrow   \rangle )
( l( \langle \rightleftharpoons m \rangle )r(  \langle n\rightleftharpoons \rangle ) )^*
l ( \langle - m \rightarrow \rangle )  \]

Right to right movement is dealt with in a similar way; the main point to note is that 
if the read head does not pass through the point $@$, then the contents of the tape to the left of the point $@$ cannot change (this is the interpretation of the set of all possible words of $\Sigma^m$ being put on the left of words of $\Sigma^n$ by the $r_n$ homomorphism). The case when the read head does pass through this point is then dealt with by a similar crossing number argument, to give 
\[ \langle m+n \rightleftharpoons \rangle =r ( \langle n \rightleftharpoons \rangle ) \cup 
r (\langle - n \rightarrow \rangle ) l(  \langle m\rightleftharpoons \rangle )  (r( \langle \rightleftharpoons n \rangle ) l( \langle m\rightleftharpoons \rangle ) )^*
r ( \langle  \leftarrow  n - \rangle ) \]
Again, duality gives the left to left movement as 
\[ \langle \rightleftharpoons m+n \rangle =r ( \langle \rightleftharpoons m \rangle ) \cup 
l ( \langle \leftarrow m - \rangle ) r( \langle \rightleftharpoons n \rangle )  (l ( \langle m \rightleftharpoons  \rangle ) r( \langle \rightleftharpoons n \rangle ) )^*
l( \langle -  m \rightarrow \rangle ) .\]
Comparing the above four terms with the vertical composition, $\circ$, of $\bf IntRel$ gives 
$\langle m+n\rangle = R_{m} (\langle n \rangle )\circ L_{n}(\langle m \rangle )$, as required. $\Box$ \\

\begin{corol}
From the computational interpretation, it is immediate that if 
\[ A\in IntRel((C_n,C_n),(C_n,C_n)) \ ,\ B\in IntRel((C_m,C_m),(C_m,C_m)) \] 
are computation relations for a bounded Turing machine ${\Bbb T}$, then 
\[ L_m(A)R_n(B)=L_n(B)R_m(A) .\]
That is, they both give the computation relations for the behavior of ${\Bbb T}$ on a tape of length $m+n$. $\Box$
\end{corol}
Although this is a necessary condition for computation relations to satisfy, it is not known whether this is a characterisation of computation relations.

\subsection{Bounded Turing machine models as monoid homomorphisms}
We demonstrate that the map $\langle \ \rangle$ can be considered to be a monoid homomorphism, as follows: \\
{\bf Definition} \\
We define 
\[ T_\infty = \bigcup_{i=0}^\infty {\bf Intrel}((C_i,C_i),(C_i,C_i)) \times \{ i\}  \]
and define a composition on $T_\infty$ by 
\[ (B,m)* (A,n) = (R_m(A)\circ L_n(B),m+n) \]
where $\circ$ is the vertical composition of ${\bf Intrel}((C_{m+n},C_{m+n}),(C_{m+n},C_{m+n}))$.
\begin{lemma}
$(T_\infty,* )$ is a monoid.
\end{lemma}
{\bf Proof} it is immediate by Lemma 7 that $*$ is associative. Also, the identity of the monoid at $(Q\times \lambda ,Q\times \lambda )$ is an identity for this composition. Our result then follows. $\Box$
\begin{theorem}
Let ${\Bbb T}$ denote a bounded Turing machine. 
The map $t:{\Bbb N}\rightarrow T_\infty$ defined by $t(n)=\langle n \rangle$ is a monoid homomorphism.
\end{theorem}
{\bf Proof} By definition, $t(\lambda)$ is the identity of $T_\infty$, and associativity follows from theorem 8. $\Box$ \\

\subsection{Extracting information from computation relations}
An objection that could be raised to the definitions of this paper is that the computation relations for a bounded Turing machine give no information about specific computations, as they describe {\em all possible} computations on a given tape length. We now demonstrate how information about specific computations can be extracted from computation relations.

Let ${\Bbb T}$ be a bounded Turing machine, and consider its computation relations on a tape of length $n$. Information about computation that go from left to right can be extracted from the relation $\langle \leftarrow n - \rangle$. Consider $q\in Q$ and $u\in \Sigma^n$. By construction, 
\[ 
\left( \left[ \begin{array}{c} q' \\ v \end{array} \right] , \left[ \begin{array}{c} q \\ u \end{array} \right] \right)  \in 
\langle \leftarrow n - \rangle 
\left\{ \left( \left[ \begin{array}{c} q \\ u \end{array} \right] , \left[ \begin{array}{c} q \\ u \end{array} \right] \right) \right\} \]
implies that 
$u^q \ \mapsto  \ ^{q'}v$ is a computation of ${\Bbb T}$.

Similarly the composition
\[  \langle - n \rightarrow \rangle 
\left\{   \left( \left[ \begin{array}{c} u \\ q \end{array} \right] , \left[ \begin{array}{c} u \\ q \end{array} \right] \right) \right\} .\]
will be the set of pairs 
$ \left( \left[ \begin{array}{c} q' \\ v \end{array} \right] , \left[ \begin{array}{c} q \\ u \end{array} \right] \right) $ 
satisfying $\ ^q u\ \mapsto  v^{q'}$ is a computation of ${\Bbb T}$. 
\\
If we restrict by an idempotent on the left hand side, instead of the right, then 
the composite 
\[ 
\left\{   \left( \left[ \begin{array}{c} q \\ u \end{array} \right] , \left[ \begin{array}{c} q \\ u \end{array} \right] \right) \right\} \langle \leftarrow n - \rangle \]
will be the set of pairs 
$\left\{ \left( \left[ \begin{array}{c} q \\ u \end{array} \right] , \left[ \begin{array}{c} q' \\ v \end{array} \right] \right) \right\}$
satisfying $\ ^{q'}v\ \mapsto \ u^q$ is a computation of ${\Bbb T}$. 
So, we can also specify the final state of a computation, and use the computation relation to calculate the set of initial states that lead to it.

We can, of course, calculate with the other computation relations in a similar way.

\section{Conclusion, and discussion of methods} 
In the above sections, we have demonstrated how symmetry ideas are useful in the basic level of theoretical computing. It is encouraging to see the emergence of the same mathematical tools that are used in both linear logic and theoretical physics. A natural point that is missing for the above is any discussion of generalisations of the syntactic monoid of an automaton
-- that is, of the languages recognised by bounded Turing machines. However, this appears to be a significantly more complicated subject; the question of whether bounded Turing machines have the same recognising power in the deterministic and non-deterministic case is still undecided (See  \cite{HU}, p.229), and a solution, whether positive or negative, appears to have important consequences. Constructing algebraic models of transitions is just the first step in answering this question.

The way in which the mathematical symmetries follow the intuitive ideas of dualising  automata is apparent in the first two cases. For the generalisation to non-deterministic automata, every relation $R$ can be written as $G^{-1}F$ for functions $F,G$. For the generalisation to two-way automata, the mathematical representation is Joyal, Street, and Verity's $\bf Int$ construction, where every object $A$ is given a dual $A^\vee$. However, there is no analogous mathematical representation for the generalisation to either Mealy machines, or bounded Turing machines. Not only that, but the monoid $T_\infty$ can be thought of as taking a copy of the endomorphism monoid of $(Q,Q)$ at {\em each} word in $\Sigma^*$ --- this is much more that a dualising process. 

In the above generalisations of finite state automata, we were (at least partially) motivated by our end-point; we already knew which models of computation we expected to construct. However, there were other possible routes to take:
\begin{itemize}
\item
For the 'reversing arrows' symmetry, the replacement of functions by relations was possibly too much of a generalisation. An alternative possibility would be to allow {\em partial injections} --- see \cite{MVL} for the resulting algebraic theory, under the name of {\em inverse  semigroups} --- and restrict them as follows: partial injections from the same state would have distinct domains, and partial injections to the same state would have distinct images (in the deterministic case, we would also require the union of the domains and of images of partial functions at a state be full). The two-way case would be slightly more complex, but the conditions required (that is, the conditions for an inverse compact closed category of partial injections) have already been found, in the context of the Geometry of Interaction, in \cite{PHD}. 
\item
For the state / alphabet symmetry, an alternative possibility would be to have the same set for the states, and the alphabet. However, although this is at first sight simpler, it means that a string of function symbols $fgh$ is ambiguous; if $f(g)=k$, and $g(h)=l$, this could denote either $k(h)$ or $f(l)$. Alternatively, it could denote the function given by applying $h$, then $g$, then $f$. Abandonment of associativity would make mathematical models significantly more complex.
\item 
It would also be reasonable to require, not only a function $\circ: \Sigma \times Q\rightarrow Q$, but also a function from $Q$ to $\Sigma \times Q$, and similarly for $*:\Sigma \times Q \rightarrow \Sigma$. Unexpectedly, I recently became aware of an application of this (in the one-way case) in the context of functional programming and automatic program transformation, in the work of Martin Erwig, \cite{ME}. 
\end{itemize}

\subsection{Acknowledgments}
I am very grateful to John Baez for several impromptu tutorials on the theory of duality and compact closure, as used in quantum mechanics, and to Ronnie Brown for introducing me to the interchange law and higher dimensional algebra. Thanks are also due to Ian France for a critical non-specialist reading of a preliminary version of this paper, and to Jon Hillier, for encouraging me to generalise two-way automata models to bounded Turing machines and pointing out some of the ways in which it was a non-trivial exercise. Finally, I would also like to thank Mark Lawson for referring me to J.-C. Birget's equations, when I was studying compact closure in a different context, and for suggesting that computation relations could be writtten in terms of a monoid, rather than a category.

\end{document}